\documentclass[12pt,epsf]{article}
\usepackage{epsfig}
\usepackage{amsmath}
\usepackage{amssymb}
\newcommand{\vp}{\varphi}

\newcommand{\LL}{{\cal L}}
\newcommand{\OO}{{\cal O}}

\newcommand{\MM}{{\cal M}}
\newcommand{\NN}{{\cal N}}

\newcommand{\PP}{{\cal P}}

\newcommand{\be}{\begin{equation}}
\newcommand{\ee}{\end{equation}}
\newcommand{\ben}{\begin{eqnarray}\displaystyle}
\newcommand{\een}{\end{eqnarray}}

\newcommand{\p}{\partial}

\newcommand{\al}{\alpha}

\newcommand{\s}{\sigma}

\newcommand{\la}{\lambda}

\newcommand{\Tr}{\hbox{Tr}}

\newcommand{\ve}{\varepsilon}

\begin{document}

{}~ \hfill\vbox{\hbox{hep-th/0503143}
\hbox{UUITP-04-05}\break
\hbox{CTP-MIT-3610}}\break

\vskip 1.cm

\centerline{\large \bf The $SU(2)$ sector in AdS/CFT}
\vspace*{5.0ex}

\centerline{\large \rm Joseph~A.~Minahan}

\vspace*{2.5ex}
\centerline{\large \it Department of Theoretical Physics}
\centerline{\large \it Box 803, SE-751 08 Uppsala, Sweden}
\vspace*{2.5ex}

\centerline{and}

\vspace*{2.5ex}
\centerline{\large \it Center for Theoretical Physics}
\centerline{\large \it Massachusetts Institute of Technology}
\centerline{\large \it Cambridge, MA 02139, USA}
\vspace*{2.5ex}

\centerline{\tt joseph.minahan@teorfys.uu.se}

\vspace*{1.0ex}
\medskip
\bigskip\bigskip
\centerline {\bf\large Abstract}

\bigskip\bigskip

In the large $N$ limit of $\NN=4$ Super Yang-Mills,
 the mixing under dilatations of  the $SU(2)$ sector, single trace
operators composed
of $L$ complex scalar fields of two types, 
is closed to all orders in perturbation theory.
By relying on the AdS/CFT correspondence, and by examining the currents for semiclassical strings, 
we present evidence which implies that
there are small mixings that contradict the closure
of the $SU(2)$ sector in the strong coupling limit.  
These mixings first appear to second
order in the 
$\la/L^2$ expansion.

\bigskip\bigskip\bigskip\bigskip
\noindent {\it To appear in proceedings of the RTN workshop,
``The quantum structure of space-time and the geometric nature of fundamental interactions'',  Kolymbari, Crete, September 2004.}

\bigskip

\vfill \eject
\baselineskip=17pt


Large $N$ gauge theories are dominated by planar diagrams, 
the diagrams 
looking like  world-sheets swept out by a  string \cite{'tHooft:1973jz}.
The string theory ``dual'' 
to a particular 4 dimensional gauge
theory  lives on a curved, higher dimensional
 manifold \cite{Polyakov:1997tj}.  The formulation of
this duality was made precise in the case of $\NN=4$ Super Yang-Mills (SYM),
where it was  argued that the string dual
is type IIB 
propagating on an $AdS_5\times S_5$  target space with
Ramond-Ramond 5-form background flux 
\cite{Maldacena:1997re,Gubser:1998bc,Witten:1998qj}.  
Among its many consequences, the AdS/CFT correspondence 
equates the dimensions of the gauge invariant operators with
the  
energies, conjugate to the global time coordinate,
 of  the dual string states in appropriate units.

In the large $N$ limit, we can restrict the gauge invariant operators 
to  be linear combinations of single trace operators.  
The problem of solving for the dimensions
of these operators can then be mapped to the problem of solving for
the energies of a one-dimensional spin chain,  the different fields in the
$PSU(2,2|4)$ singleton multiplet corresponding to the possible ``spins'' at
each site \cite{Minahan:2002ve,Beisert:2003jj,Beisert:2003yb}. The hamiltonian for the spin chain is the dilatation operator
of the full superconformal algebra.

In order to simplify the problem, it is convenient to look for closed
sectors of $PSU(2,2|4)$.  The simplest nontrivial sector is the $SU(2)$ 
sector, where the operators are restricted to contain
only two types of complex
scalar fields, say $Z$ and $W$.  For a massless theory, dilatations
only mix operators with the same bare dimension.  So if one considers an operator $\OO$
that is a single trace composed of $J_1$ $Z$  fields and
$J_2$ $W$ fields, then it is straightforward to show using
the $R$-symmetry and Lorentz invariance that 
dilatations  can only  mix  $\OO$ with  operators
which have the same number
of $Z$ and
$W$ fields {\it to all orders in perturbation theory}.  
In the $SU(2)$ sector, the one loop  dilatation
operator is equivalent to the Heisenberg spin chain with nearest neighbor
interactions.  At $n$ loop order, the interactions between the spins
range over $n$ sites, and   the dilatation operator restricted to this
sector is  
known, with a few assumptions, to five loop order and
is consistent with integrability 
\cite{Beisert:2003tq,Beisert:2003ys,Beisert:2004ry}, 

In order to test the AdS/CFT duality, one needs to find the eigenvalues
of the dilatation operator acting on  operators with $L=J_1+J_2$
total fields, and then compare them to the energies of the corresponding
string states.  This program has met with apparent success in the BMN
limit \cite{Berenstein:2002jq,Gross:2002su,Santambrogio:2002sb}, 
where the operators are very close to being BPS.  It has also
been successful, to a point,
 in the long wave-length limit, where the eigenvalues
of the dilatation operator have been shown to match the string predictions
at the one-loop level \cite{Beisert:2003xu,Frolov:2003xy,Beisert:2003ea,Arutyunov:2003rg,Engquist:2003rn,Kruczenski:2003gt,Mikhailov:2003gq,Engquist:2004bx,Kristjansen:2004ei,Kazakov:2004qf,Hernandez:2004uw,Stefanski:2004cw,Mikhailov:2004xw,Smedback:1998yn,Freyhult:2004iq,Kristjansen:2004za,Kruczenski:2004cn,Mikhailov:2004au,Hernandez:2004kr,Kazakov:2004nh,Beisert:2004ag,Schafer-Nameki:2004ik,Beisert:2005bm,Frolov:2004bh,Park:2005ji,Beisert:2005mq}, 
and in some cases 
up to two loop order 
\cite{Serban:2004jf,Kazakov:2004qf,Kruczenski:2004kw,Minahan:2004ds,Staudacher:2004tk}, although there is a mismatch
at three loop order 
\cite{Callan:2003xr,Serban:2004jf,Beisert:2004hm,Minahan:2004ds}.  
Both the BMN limit and the long wave-length
limit can be treated using a semiclassical analysis \cite{Gubser:2002tv,Frolov:2002av}, with expansion parameter  $\lambda/L^2$, where $\lambda$ is the 
't Hooft parameter and quantum corrections are suppressed by $1/L$.

In any case, we can now ask the following question: is there a closed
$SU(2)$ sector for the dual string?\,\footnote{
My talk at the RTN conference
was a review of the use of spin chains to compute anomalous dimensions
in SYM, with the goal of comparing these results to
string theory.  Since there are already excellent published reviews on this
subject \cite{Tseytlin:2003ii,Beisert:2004ry}, and since many of the questions were about the three loop disagreement
between gauge theory and string theory predictions, I have decided
to present in these precedings some previously unpublished work about
another comparision between SYM and semiclassical string results.}  
The closure of the $SU(2)$ sector
in $\NN=4$ SYM is shown at the perturbative level.  However, computations
for the dual string are best under control when the coupling is large,
where perturbative statements can break down.

In SYM, the dilatation operator becomes more and more complicated as we
go to higher orders in perturbation theory, and it appears unlikely
that an all orders expression will be found \footnote{
Although it might be possible to find the S-matrix for
the spin chain to all orders \cite{Staudacher:2004tk}.}.  
Nevertheless, we do
know that there are approximately
$2^L/L$
independent single trace $SU(2)$ sector operators in the large $N$ limit\footnote{
For the spin chain there are $2^L$ states, but for the gauge theory
there is a trace condition that reduces this number somewhat.}.
If one could establish a closed $SU(2)$ sector on the string side, 
then 
one would also like to see the $2^L/L$ string states that are dual
to these operators.  
Clearly most of these states are outside the long wave-length
limit, so a full quantization of the string theory is required.

Let us now examine the question of $SU(2)$ closure in the dual string picture,
considering semiclassical states that are dual
to long wavelength operators.  Since the operators dual to the strings have
only
two types of scalar fields and no covariant derivatives, the semiclassical
string motion is constrained to an $R\times S_3$ subspace of the full
$AdS_5\times S_5$.  The isometry group of $S_3$ is $SO(4)=SU(2)_L\times SU(2)_R$ where we have labeled the $SU(2)$ subgroups in terms of left and right
currents on the $SU(2)$ sigma model.  We can express the coordinates
on $S_3$ in terms of an $SU(2)$ group element
\begin{equation}
g=\left(\begin{array}{cc}
Z&W\\
-\overline W&\overline Z
\end{array}\right),
\end{equation}
where
\begin{equation}\label{norm}
|Z|^2+|W|^2=1.
\end{equation}
The sigma model action is invariant under two global $SU(2)$ transformations,
$g(\tau,\s)\to g(\tau,\s)U$ and $g(\tau,\s)\to Vg(\tau,\s)$.
Hence, we can construct a right  current, 
{$j_\al=-ig^{-1}\p_\al g$,} and
a left  current, $\ell_\al=-i\p_\al gg^{-1}$.

If we treat the sigma model classically, then it is convenient to
treat $j_1^a(\s)$ and $\ell_1^a(\s)$ as canonical coordinates 
\cite{Faddeev:1987ph}, 
whose Poisson
brackets then satisfy\footnote{
In \cite{Faddeev:1985qu} an alternative set of Poisson 
relations has been proposed, which necessarily uses a different Hamiltonian
for the time evolution of the currents.  
This Hamiltonian naturally
appears as one of the local charges in the sigma model \cite{Zarembo:2004hp}.
I thank K. Zarembo for remarks on this point.}
\begin{equation}
\{j_1^a(\s),j_1^b(\s')\}=\{\ell_1^a(\s),\ell_1^b(\s')\}=\{j_1^a(\s),\ell_1^b(\s')\}=0.
\end{equation}
The time-like components of the currents are not precisely the canonical
momenta, but instead have Poisson brackets given by
\begin{eqnarray}
\{j_0^a(\s),j_1^b(\s')\}&=&i\ve^{abc}j_1^c(\s)\delta(\s-\s')+\delta^{ab}\delta'(\s-\s')\nonumber\\
\{\ell_0^a(\s),\ell_1^b(\s')\}&=&i\ve^{abc}\ell_1^c(\s)\delta(\s-\s')+\delta^{ab}\delta'(\s-\s')\nonumber\\
\{j_0^a(\s),\ell_1^b(\s')\}&=&ig^{-1}(\s)t^ag(\s')\delta'(\s,\s'),
\end{eqnarray}
and
\begin{eqnarray}\label{tlcs}
\{j_0^a(\s),j_0^b(\s')\}&=&i\ve^{abc}j_0^c(\s)\delta(\s-\s')\nonumber\\
\{\ell_0^a(\s),\ell_0^b(\s')\}&=&i\ve^{abc}\ell_0^c(\s)\delta(\s-\s')\nonumber\\
\{j_0^a(\s),\ell_0^b(\s')\}&=&0\,,
\end{eqnarray}
where $t^a$ are $SU(2)$ generators satisfying 
\begin{equation}
[t^a,t^b]=i\ve^{abc}t^c,\qquad\qquad \Tr (t^at^b)=\frac12\delta^{ab}.
\end{equation}
Notice that the left and right currents are not completely 
independent.

{}From the Poisson brackets in (\ref{tlcs}), we can  start to see why the 
spin chain is related to the sigma model \cite{Kruczenski:2003gt}.  
Suppose that we have a one dimensional chain with $L$ sites and two 
independent spin operators at each site $k$, 
$\vec S_k$ and $\vec T_k$, satisfying the commutation relations
\begin{equation}\label{commrel}
[S^a_k,S^b_{k'}]=i\ve^{abc}S^c_k\delta_{kk'}\,,\qquad
[T^a_k,T^b_{k'}]=i\ve^{abc}T^c_k\delta_{kk'}\,,\qquad
[S^a_k,T^b_{k'}]=0.
\end{equation}
In the limit that $L\to\infty$, the commutation relations in (\ref{commrel})
would match
the quantized version of (\ref{tlcs}), where
the Poisson brackets are replaced by Dirac brackets.  Or put another
way, we can identify the expectation values of $S^a_k$ and $T^a_k$
with the classical currents as 
\begin{equation}\label{spinmatch}
\langle S^a_k\rangle=j^a_0\left(2\pi k/L\right),\qquad\qquad \langle T^a_k\rangle=
\ell^a_0\left({2\pi k/L}\right).
\end{equation}

Returning to the components of $g$, $(Z,W)$ and $(-\overline W,\overline Z)$ 
transform as doublets under  
$SU(2)_R$ transformations, while $(Z,-\overline W)$ and $(W,\overline Z)$ 
transform as doublets under  $SU(2)_L$ transformations.  Hence, 
for operators in
the $SU(2)$ sector, the $SU(2)_L$ spin is up at every site.
This suggests that the desired classical solutions  should
have a left current satisfying
\begin{equation}\label{highest}
\ell_0= \left(\begin{array}{cc}
\ell&0\\
 0&-\ell
\end{array}\right)\, .
\end{equation}
where $\ell$ is assumed to be a function of $\tau$ and $\sigma$ \footnote{
The $\tau$ and $\sigma$ dependence is gauge dependent, but under an
appropriate gauge choice, $\ell$ can be set to a constant $L$ 
\cite{Kruczenski:2004kw}.}.  
This would correspond to all  left charges aligning in the same
direction, indicating that the dual operator only has $Z$ and
$W$ fields and not their complex conjugates.

But the condition in (\ref{highest}) is overly
restrictive.  The condition that $\ell_0$ have no off-diagonal
terms immediately leads to the relation
$Z=f(\s)W$.  This, along with (\ref{norm}), means that
\begin{equation}
\p_0(|Z|^2)=\p_0(|W|^2)=0.
\end{equation}
Hence, we can write $Z=g_1(\s)e^{iw(\tau)}$, $W=g_2(\s)e^{iw(\tau)}$,
with
\begin{equation}
|g_1(\s)|^2+|g_2(\s)|^2=1.
\end{equation}

The relevant sigma model action in conformal gauge is
\begin{equation}\label{action}
S=\frac{\sqrt{\lambda}}{2\pi}\int d\tau\int_0^{2\pi}d\s\frac{1}{2}
\left(\frac12(\Tr(\ell_0\ell_0)-\Tr(\ell_1\ell_1))-\p_\al t\p^\al t\right)
\end{equation}
{}From the reparameterization invariance we can
choose $t=\kappa\tau$, where $\kappa$ is a constant, which leads to
the constraint
\begin{equation}\label{energyrel}
\kappa^2=\frac{1}{2}\Big(\Tr(\ell_0^2) +\Tr(\ell_1^2)\Big)=
\left(\frac{dw}{d\tau}\right)^2
+|\bar g_1 g_1'+\bar g_2 g_2'|^2+|g_1g_2'-g_2g_1'|^2.
\end{equation}
Since $g_1, g_2$ and $\kappa$ are independent of $\tau$, we see that
$w=w_0\tau$ and so $\ell_0$ is also independent of $\tau$.  
The other constraint is
\begin{equation}
\Tr(\ell_0\ell_1)=0,
\end{equation}
which gives
\begin{equation}\label{eq1}
\bar g_1 g_1'+\bar g_2g_2'=0.
\end{equation}
Finally the equation of motion is
\begin{equation}
\p_0\ell_0-\p_1\ell_1=0.
\end{equation}
Since $\ell_0$ is independent of $\tau$, we have
$\p_1\ell_1=0$, which leads to the relation
\begin{equation}\label{eq2}
g_1g_2''=g_2g_1''.
\end{equation}

Eqs. (\ref{eq1}) and (\ref{eq2}) can be satisfied if either $g_1$ or
$g_2$ is zero.  Such a solution would correspond to a point like string
which is dual to the chiral primary operator $\Tr(Z^L)$.  If neither $g_1$ nor $g_2$
is zero, then it is convenient to write these functions as
$g_j=\rho_je^{i\theta_j}$ where $\rho_j$ and $\theta_j$ are real functions
of $\s$, and so
\begin{equation}
\rho_1^2+\rho_2^2=1.
\end{equation}
  Then from (\ref{eq1}) we find that
\begin{equation}\label{eq3}
\rho_1^2\theta_1'=-\rho_2^2\theta_2',
\end{equation}
and from (\ref{eq2}) we also find
\begin{equation}\label{eq4}
(\rho_1^2\theta_1')'=(\rho_2^2\theta_2')'=0.
\end{equation}
Plugging (\ref{eq3}) and (\ref{eq4}) into (\ref{energyrel})
gives the general solution
\begin{equation}
\rho_1^2=\frac{1}{2}(1+A\cos (2m\s+\delta)),
\end{equation}
with
\begin{equation}
\kappa^2=w_0^2+m^2\,,
\end{equation}
where $m$ is an integer and $-1<A<1$.
Hence, these solutions correspond to the Frolov and Tseytlin circular
string solutions with $J_1=J_2$ \cite{Frolov:2003qc}.  
In fact, the constant $A$ can be set
to zero
by choosing a new basis for $Z$ and $W$.
Hence, except for the point-like  and the Frolov-Tseytlin circular string,
it is not possible to find a classical solution where the $SU(2)_L$ charge
is aligned, that is proportional to $\s_3$, everywhere on the string.

It is instructive to write the left currents in terms of
angles on $S_3$.  If we express $Z$ and $W$ as 
\begin{equation}
Z=\cos\theta e^{i\phi_1}\qquad W=\sin\theta e^{i\phi_2}\,,
\end{equation}
then the left current can be written as
\begin{equation}\label{leftmat}
\ell_0=\left(\begin{array}{cc}
\dot\vp_1+\dot\vp_2\cos2\theta&(-i\dot\theta+\dot\vp_2\sin2\theta)e^{2i\vp_1}\\
 (i\dot\theta+\dot\vp_2\sin2\theta)e^{-2i\vp_1}&-\dot\vp_1-\dot\vp_2\cos2\theta
\end{array}\right),
\end{equation}
where $\vp_{1,2}=\frac{1}{2}(\phi_1\pm\phi_2)$.
The condition that $\ell_0$ have no off-diagonal terms leads to 
$\dot\theta=\dot\vp_2=0$, which are precisely the conditions found
by Kruczenski for the restriction of the $\sigma$-model to match the
spin chain effective action at one-loop \cite{Kruczenski:2003gt}.  
However, because of the
constraints, these restrictions on $\dot\theta$ and $\dot\vp_2$ do
not hold beyond the one-loop level.

To see what happens beyond the one-loop level, let us
 consider a particular class of solutions where the charge is
uniformly distributed on the string in conformal gauge 
\cite{Kruczenski:2004kw}.  For these solutions
we choose the ansatz \cite{Arutyunov:2003uj,Arutyunov:2003za}
\begin{equation}
Z=\cos\theta_0e^{iw_1\tau+m_1\s}\qquad\qquad W=\sin\theta_0 e^{iw_2\tau+m_2\s}.
\end{equation}
The constraints lead to the equations 
\begin{eqnarray}
\kappa^2&=&\cos^2\theta (w_1^2+m_1^2)+\sin^2\theta(w_2^2+m_2)^2\nonumber\\
0&=&w_1m_1\cos^2\theta+w_2m_2\sin^2\theta
\end{eqnarray}
and the equation of motion leads to
\begin{equation}
w_1^2-m_1^2=w_2^2-m_2^2.
\end{equation}
Substituting into (\ref{leftmat}), we find
\begin{equation}\label{leftmat2}
\ell_0=\left(\begin{array}{cc}
\frac{1}{2}(w_1+w_2)+\frac{1}{2}(w_1-w_2)\cos2\theta_0&\frac{1}{2}(w_1-w_2)\sin2\theta_0e^{i(w_1+w_2)\tau+i(m_1+m_2)\s}\\
 \frac{1}{2}(w_1-w_2)\sin2\theta_0e^{-i(w_1+w_2)\tau-i(m_1+m_2)\s}&-\frac{1}{2}(w_1+w_2)-\frac{1}{2}(w_1-w_2)\cos2\theta_0
\end{array}\right),
\end{equation}
From the equations of motion, we see that $w_1-w_2\approx (m_1^2-m_2^2)/2w_1$
and $w_1\approx L/\sqrt{\lambda}$.  Hence, we see that $\ell_0=\vec\ell_0\cdot
\vec\s$ where $|\ell^{1,2}_0|\sim \frac{\lambda}{L^2}|\ell^3_0|$.  In
other words, there is a small component of the left current
that is not aligned in the $\hat z$ direction\footnote{
We can also see that the the solution in (\ref{leftmat2}) is of
Landau-Lifschytz type.  If we write $\ell_0=\vec T\cdot\vec \s$, then
we see that to two loop order, the components of $\vec T$ satisfy
\begin{equation}
\dot{\vec T}\approx \frac{1}{(m_1+m_2)^2} \vec T\times \vec T''\, .
\end{equation}
It is not clear what the significance of this equation is,
especially the meaning of the coefficient that depends on
the winding.}.  
It is clear that this
is a two loop effect, 
since the contribution to $\kappa^2$ from this
term is $|\ell_0^1|^2+|\ell_0^2|^2$.  
One can also show that the nondiagonal piece cannot be removed by 
choosing a different gauge for the Polyakov action, such as the nondiagonal
gauge used in \cite{Kruczenski:2004kw}.

One way to interpret these results is that there is mixing
outside of the $SU(2)$ sector that is nonperturbative in $\lambda$,
with the string capturing this nonperturbative behavior.   This
could occur through a 
 reversal in the order of limits, where on the string side one first takes
$L\to\infty$  and then expands in $\lambda/L^2$, while in the
gauge theory, one first expands in $\lambda$ and then takes $L\to\infty$
\cite{Serban:2004jf,Kazakov:2004qf,Beisert:2004hm}, although I know
of no specific process through which this could occur\footnote{
I thank M. Staudacher for remarks on this point.}.

The nondiagonal component of the left current
 also leads to an uncertainty in what is meant
by the bare dimension of the operator.  In comparing string solutions
to Yang-Mills solutions, one usually assumes that the
bare dimension is the left charge, since on the Yang-Mills side there
are only $Z$ and $W$ scalar fields.  However, if there are also $\bar Z$
and $\bar W$ fields, then the bare dimension would be larger than $L$,
since $L$ is measuring the net number of unbarred fields minus the
number of barred fields.  If the bare dimension is larger, then this
means that the anomalous dimension is smaller at the two loop level
and so there would be a mismatch between the gauge theory 
and string theory predictions.  

It is also possible that the $SU(2)$ sector {\it is} closed at strong
coupling, but there is a subtlety in identifying the left and right currents
with the spins on the chain as in (\ref{spinmatch}).  However, the
identification in (\ref{spinmatch}) seems quite natural and there
is no obvious alternative.

We can also examine the possible mixing by looking at the finite
gap solutions of the principle chiral model.
In \cite{Kazakov:2004qf}
a one to one map was found between the long wave-length solutions
of the $SU(2)$ sector and the corresponding solutions for the sigma
model.  This work was later extended into other sectors
\cite{Kazakov:2004nh,Beisert:2004ag,Schafer-Nameki:2004ik,Beisert:2005bm}.  
The relevant facts can be found in \cite{Kazakov:2004qf}, 
but on the SYM side,
the one-loop Bethe equations can be reduced to
solving for a resolvant, 
\begin{equation}
G(\vp)=\int\frac{d\vp'\rho(\vp')}{\vp-\vp'}\,,
\end{equation}
where $\rho(\vp)$ is the density of Bethe roots in the complex
plane of the spectral parameter $\vp$.   $G(\vp)$ has square root branch
cuts and  asymptotic behavior
\begin{equation}
G(\vp)\sim \frac{J_2/L}{\vp}\qquad\qquad \vp\to\infty.
\end{equation}
$G(\vp)$ is also the generator of the conserved charges 
\cite{Arutyunov:2003rg,Engquist:2003rn}, with
$G(0)=-2\pi m$ being the total momentum of the impurities in the spin
chain, and $G'(0)=8\pi^2L\gamma/\lambda$ giving the anomalous dimension.

In the long wave-length limit, it is also possible to consider
larger classes of scalar operators that live in an $SO(4)$ sector.
Strictly speaking, this sector does not close, even at one-loop, but
in the long-wave length limit, this nonclosure is $1/L$ effect and can
be ignored \cite{Minahan:2004ds}.  
These $SO(4)$ solutions can then be considered
as two independent $SU(2)$ chains, with the resolvant being the
sum of the two, $G(\vp)=G_1(\vp)+G_2(\vp)$ where $G_{1,2}(\vp)$ is the resolvant
for each $SU(2)$.  Since each $G_i(\vp)$ is a generator of conserved
charges, we could also consider $\tilde G(\vp)=G_1(\vp)-G_2(\vp)$.  However,
the anomalous dimension is in $G(\vp)$ so this is the one we are interested
in.

Higher orders in $\lambda$ will modify the equations that the $G_i(x)$
satisfy, but in the long wave-length limit, we still expect the 
$SU(2)$ sectors to remain separated.  Hence, the various solutions can be
written as functions on a 4 sheeted surface, with branch cuts connecting
the first surface to the second  and the third surface to the
fourth.

On the string side, there is an analog of $G(x)$.  The $\sigma$-model
on $R\times S_3$ is classically integrable \cite{Pohlmeyer:1975nb}\footnote{
The $\sigma$-model on the full $AdS_5\times S_5$ is also 
classically integrable \cite{Mandal:2002fs,Bena:2003wd,Dolan:2003uh,Arutyunov:2003uj,Arutyunov:2003za,Alday:2003zb,Swanson:2004mk,Arutyunov:2004yx,Alday:2005gi}.}. 
This means that there is
a lax pair
\begin{eqnarray}
\LL&=&\p_\s-\frac{i\sqrt{T}}{2}\left(\frac{j_+}{x-\sqrt{T}}+\frac{j_-}{x+\sqrt{T}}\right)\nonumber\\
\MM&=&\p_\tau-\frac{i\sqrt{T}}{2}\left(\frac{j_+}{x-\sqrt{T}}-\frac{j_-}{x+\sqrt{T}}\right)\nonumber
\end{eqnarray}
that satisfies the flatness condition
$[\LL,\MM]=0$, which is a consequence of the equation of
motion
\begin{equation}
\p_+j_-+\p_-j_+=0
\end{equation}
and the identity
\begin{equation}
\p_+j_--\p_-j_++i[j_+,j_-]=0.
\end{equation}
The parameter $T$ is $\frac{\lambda}{16\pi^2L^2}$ and $x$ is a spectral
parameter.
We can then look for the solution of the equation
\begin{equation}
\LL\psi=0,
\end{equation}
which is formally given by the path ordered expression
\begin{equation}
\psi(\s,x,\tau)=\PP\exp\left(\frac{i\sqrt{T}}{2}\int_0^\s d\s'\left(\frac{j_+}{x-\sqrt{T}}+\frac{j_-}{x+\sqrt{T}}\right)\right).
\end{equation}
Setting $\Omega(x)=\psi(2\pi,x,\tau)$, it is immediately clear that $\Omega$
is constant in $\tau$ because of the flatness condition.  Under a similarity
transformation, $\Omega(x)$ can be diagonalized to
\begin{equation}
\Omega(x)=\left(\begin{array}{cc}e^{ip(x)}&0\\0&e^{-ip(x)}\end{array}\right)\, ,
\end{equation}
where $p(x)$ can have square root branch cuts,  the branch points occurring
at values of $x$ where $e^{ip(x)}=\pm1$.  

Asymptotically, $p(x)$ behaves as
\begin{equation}
p(x)\sim-\frac{\pi\kappa\sqrt{T}}{x\mp\sqrt{T}}\qquad x\to\pm\sqrt{T}
\end{equation}
\begin{equation}\label{pxinfty}
p(x)\sim -\frac{ Q_R}{2Lx}\qquad x\to\infty
\end{equation}
and
\begin{equation}\label{pxzero}
p(x)=2\pi m +\frac{Q_L}{2LT}x+\ldots\qquad x\to0,
\end{equation}
where $Q_{R,L}$ are the left and right charges
\begin{eqnarray}
Q_{R}&=&\frac{\sqrt{\la}}{2\pi}\int_0^{2\pi}j_0^3 d\s\nonumber\\
Q_{L}&=&\frac{\sqrt{\la}}{2\pi}\int_0^{2\pi}\ell_0^3 d\s\, .
\end{eqnarray}
We can then define  $G_s(x)=p(x)+\frac{(\Delta/L) x}{2(x^2-T)}$
which has no poles on one of the sheets.  
$G_s(x)$ has square root
branch cuts on a two-sheeted surface, but it has information about both
$SU(2)_R$ and $SU(2)_L$.  If all branch points are at $|x|>>\sqrt{T}$ as
$T\to0$, then it was shown in \cite{Kazakov:2004qf} that $G_s(x)$ approaches $G_1(x)$.  
But it was also shown in \cite{Beisert:2004ag} that there is an inversion symmetry,
such that $G(x)\to -G(T/x)+2\pi n$ switches $SU(2)_R$ with $SU(2)_L$.    
Hence,
if all branch points are located inside $|x|<<\sqrt{T}$, then
$-G_s(T/x)+2\pi n$ approaches $G_2(x)$.  However, if there are branch points
at both extremes, then we should consider the total contribution, and
so we would have
\begin{equation}
G(x)=G_1(x)+G_2(x)=G_s(x)-G_s(T/x)+2\pi n\qquad T\to0.
\end{equation}

Two simple examples were given in \cite{Kazakov:2004qf}.  The first corresponds
to the Frolov-Tseytlin solution with $J_1=J_2$.  In this case
\begin{equation}
G_s(x)=\frac{1}{2}\frac{x}{x^2-T}\left(\Delta/L+\sqrt{16\pi^2m^2x^2+1}\right)
-2\pi m,
\end{equation}
where $m$ is the winding and $\Delta=\sqrt{L^2+m^2\la}$.
Hence, 
\begin{equation}
G_1(x)=\frac{1}{2x}\left(1+\sqrt{16\pi^2m^2x^2+1}\right)-2\pi m\,,
\end{equation}
 and $G_2(x)=0$ as $T\to0$.  
Note that the separation into the $SU(2)_R$ or $SU(2)_L$ sectors
breaks down when $m^2\sim L^2/\lambda$, which is also where the
semiclassical string approximation breaks
down.  On the SYM side, this corresponds
to the case where the anomalous dimension is of order $L$, the
same order as the bare dimension.

The second example is for a pulsating string, where $G_s(x)$ is
\begin{equation}
G_s(x)=\frac{1}{2}\frac{x}{x^2-T}\left(\Delta/L+\sqrt{\left(2\pi m(x-T/x)-
\frac{J}{L}\right)^2+\frac{\Delta^2-J^2}{L^2}}\right)-\pi m,
\end{equation}
where $J=Q_R=Q_L$.   Hence, we find that
\begin{eqnarray}
G_1(x)&=&\frac{1}{2x}\left(1+\sqrt{(2\pi mx-J/L)^2+1-J^2/L^2}\right)-\pi m
\nonumber\\
G_2(x)&=&\frac{1}{2x}\left(1+\sqrt{(2\pi mx+J/L)^2+1-J^2/L^2}\right)+\pi m
\end{eqnarray}
as $T\to0$.

It is hence clear that the left and right pieces have effectively separated
in the limit $T\to0$.  Essentially we have taken our two sheeted surface
and cut a small hole of radius $\sqrt{T}$ out of each sheet.  In the limit
that $T\to 0$, the disc and the sheet with the hole in it 
separate into two sheets with no cuts connecting them\footnote{
We are assuming that there are an even number of branchpoints both inside
and outside the radius $\sqrt{T}$ so that there are no cuts running between the
two regions.  This is required in order to match to perturbative SYM.
I thank N. Beisert for pointing out an incorrect statement on this point in
the previous version.}.  Hence, our two-sheeted
surface has separated into four sheets, with cuts only connecting the
separate pairs. This can also be seen in the context of the full
$SO(6)$ symmetry of the $S_5$ \cite{Beisert:2004ag}, where one starts
with 4 sheets separated into pairs.  The pairs are related by the
inversion symmetry,  the  inversion taking
 the branch points inside $x=\sqrt{T}$ for one pair to the branch
points outside $x=\sqrt{T}$ for the other pair.

Once we start considering higher order corrections in $T$, which correspond
to higher orders in the Yang-Mills coupling, the separation can no longer
be maintained. To see this, note that it was possible to successfully
match all Frolov-Tseytlin solutions in string theory to SYM 
solutions up to two loop order in the $SU(2)$ sector \cite{Kazakov:2004qf},
if one assumes that the left hand charge $L$ is the bare dimension of
the SYM operators.  
In order to demonstrate the
matching, it was necessary to redefine the spectral parameter 
\cite{Kazakov:2004qf}.  
In \cite{Beisert:2004hm,Marshakov:2004py} it was
shown how to do this in a very nice fashion.  
The idea is as follows:  let us call the spectral parameter for the gauge
theory $\vp$ and the spectral parameter  for the string theory $x$.  At the one-loop level, we
have $\vp=x$.  Using (\ref{pxinfty}), (\ref{pxzero})
 and the definition of $G_s(x)$, we find
that $G_s(x)$ has the asymptotic behavior
\begin{eqnarray}
G_s(x)&\sim& \frac{J}{Lx}+\frac{\Delta-L}{2Lx},\qquad x\to \infty\nonumber\\
G_s(x)&=&2\pi m-\frac{\Delta-L}{2LT}x+\ldots\qquad x\to0,
\end{eqnarray}
where we have used that $Q_L=L$ and $Q_R=L-2J$.
We now assume that the string resolvant $G_s(x)$ can be written as 
an integral over a density of roots, $\rho_s(x)$, as
\begin{equation}
G_s(x)=\int \frac{dx'\rho_s(x')}{x-x'}
\end{equation}
and so we have
\begin{eqnarray}
\int dx'\rho_s(x')=\frac{J}{L}+\frac{\Delta-L}{2L}\nonumber\\
T\int \frac{dx'\rho_s(x')}{x'^2}=\frac{\Delta-L}{2L}.
\end{eqnarray}
For the gauge theory, we expect
\begin{equation}
\int d\vp'\rho(\vp')=\frac{J}{L},
\end{equation}
and so we can make the identification $\vp=x+T/x$ and $\rho_s(x)=\rho_1(\vp)$
\cite{Beisert:2004hm}.
Now let us see what this means for the relations between the resolvants.
We have that \cite{Beisert:2004ag}
\begin{eqnarray}\label{G1Gs}
G_1(\vp)&=&\int\frac{d\vp'\rho_1(\vp')}{\vp-\vp'}=
\int\frac{dx'(1-T/x'^2)\rho_s(x')}{(x-x')(1-T/xx')}\nonumber\\
&=&\int dx'\rho_s(x')\left(\frac{1}{x-x'}+\frac{1}{T/x-x'}+\frac{1}{x'}\right)
\nonumber\\
&=&G_s(x)+G_s(T/x)-G_s(0),
\end{eqnarray}
where $x=\frac{1}{2}(\vp+\sqrt{\vp^2-4T})$ \footnote{Note that (\ref{G1Gs})
gives the same result if we choose
the other solution, $x=\frac{1}{2}(\vp-\sqrt{\vp^2-4T})$,
since this is $T/x$ of the original solution.}.  The result in (\ref{G1Gs})
was shown to match the Yang-Mills result at the two loop level, but fails
at three loops.  

But we can also see a problem with (\ref{G1Gs}) at the two loop level.  If
there are branch points in the region $|x|<<\sqrt{T}$, then we would
want to use
$G(\vp)=G_s(x)-G_s(T/x)+2\pi n$ in order to match the {\it one-loop} result for
the full $SU(2)_L\times SU(2)_R$ theory.  Presumably, we can smoothly
adjust the solution such that the branch points in $|x|<<\sqrt{T}$
disappear, reducing everything to a single $SU(2)$.  But then we see
that we have the opposite sign for $G_s(T/x)$ as compared to (\ref{G1Gs}).
If we went further with this definition of $G(\vp)$, then we would
find that in the limit of small $T$, $G(\vp)$ would have a pole at
$\vp=0$ whose residue is
linear in $T$, and hence $G(\vp)$ would not match the gauge
theory result at this order.
So from this point of view, the two-loop matching condition seems
unnatural.  

Nevertheless, a prescription has been given to match an $SU(2)_L\times SU(2)_R$
solution at two loops \cite{Minahan:2004ds}, at least such that the dimension of
the operator matches the energy of the string.  We do not yet know
how to show matching of {\it all} conserved quantities, although
perhaps this can be accomplished with a different relation between
$\vp$ and $x$.  

In conclusion, 
the viewpoint that one should  probably take, assuming that the semiclassical
string results are trustworthy, is that it
is not sensible to split up the dimension of an operator into
a ``bare'' and an anomalous piece at strong coupling. 
If it were possible, then one
could infer that there are a fixed number of fields, which would effectively
keep the $SU(2)$ sector closed.  Instead, only the full dimension is
well defined and it is a function of $\sqrt{\lambda}$ and the charge $L$.
In going from weak to strong coupling, the long wave-length
operators in the $SU(2)$ sector
smoothly flow to the Frolov-Tseytlin solutions, such that the first
two orders in $\la/L^2$ agree.  

As one gets away from the long wavelength
limit,  it becomes harder to identify the string states which will smoothly
flow to operators in the $SU(2)$ sector as $\la$ goes from
strong to weak coupling. It may be that 
counting the $2^L/L$ string states is possible
 only if $\la$ is small, where of course
the string is strongly coupled.
A proposal has been made to
discretize the string Bethe equations, and use this as a quantized Bethe
ansatz for the string states that flow from the $SU(2)$ sector 
\cite{Arutyunov:2004vx}.  
Perhaps,
one can count the solutions of these discretized Bethe equations 
by counting all Bethe strings of this model, as was done for
the Heisenberg chain \cite{Faddeev:1996iy}.  

\bigskip

\noindent {\bf Acknowledgments}:
I thank N. Beisert, A. Marshakov, M. Staudacher and K. Zarembo for 
discussions and comments on the manuscript.  I also thank the
organizers of the RTN workshop in Kolymbari, Crete for the
invitation to speak.
This research was
supported in part by Vetenskapsr\aa det
and by DOE contract \#DE-FC02-94ER40818.

\end{document}